\documentclass[conference,a4paper]{IEEEtran}
\pdfoutput=1

\usepackage{etex}
\usepackage{pgfplots}
\pgfplotsset{compat=newest}
\usepackage{tikz}
\usetikzlibrary{arrows,matrix,positioning}
\usepackage[utf8]{inputenc}
\usepackage[innermargin=0.545in,outermargin=0.65in,top=0.545in,bottom=.7in]{geometry}
\usepackage[english]{babel}
\usepackage[T1]{fontenc}
\usepackage{epsfig}
\usepackage{amsmath, amssymb, amsbsy}
\usepackage{mathdots}
\usepackage{xspace}
\usepackage[noend]{algpseudocode}
\usepackage{algorithm}
\usepackage{algorithmicx}
\usepackage{color}
\usepackage{cite}
\usepackage{booktabs}
\usepackage{verbatim}
\usepackage{lipsum}
\usepackage{enumitem}
\usepackage{colortbl}
\usepackage{theorem}
\usepackage{flushend}

\newtheorem{theorem}{Theorem}
\newtheorem{lemma}[theorem]{Lemma}
\newtheorem{corollary}[theorem]{Corollary}

\usepackage[final,tracking=true,kerning=true,spacing=true,factor=1100,stretch=10,shrink=20]{microtype}

\renewcommand{\vec}[1]{\ensuremath{\mathbf{#1}}}

\newcommand{\0}{\ensuremath{\mathbf{0}}}

\newcommand{\Mooremat}[3]{\ensuremath{\mathcal{M}_{#1,#2}\left( #3 \right)}}
\newcommand{\MoormatExplicit}[3]{
	\begin{pmatrix}
		#1_{1} & #1_{2} & \dots& #1_{#3}\\
		#1_{1}^{q} & #1_{2}^{q} & \dots& #1_{#3}^{q}\\
		\vdots &\vdots&\ddots& \vdots\\
		#1_{1}^{q^{#2-1}} & #1_{2}^{q^{#2-1}} & \dots& #1_{#3}^{q^{#2-1}}\\
	\end{pmatrix}}
	
\newcommand{\Fq}{\ensuremath{\mathbb{F}_q}}
\newcommand{\Fqm}{\ensuremath{\mathbb{F}_{q^m}}}
\newcommand{\Fqmu}{\ensuremath{\mathbb{F}_{q^{mu}}}}

\DeclareMathOperator{\rank}{rank}
\DeclareMathOperator{\Tr}{Tr}
\newcommand{\nkhalffrac}{\left\lfloor \frac{n-k}{2}\right\rfloor}

\newcommand{\RkError}{\ensuremath{\varphi}}

\newcommand{\tpub}{\ensuremath{t_{\mathsf{pub}}}}
\newcommand{\kpub}{\ensuremath{\mathbf{k}_{\mathsf{pub}}}}
\newcommand{\mycode}[1]{\ensuremath{\mathcal{#1}}}
\newcommand{\fontmetric}[1]{\mathsf{#1}}

\newcommand{\codelinearRank}[1]{\ensuremath{[#1]_q^\fontmetric{R}}}

\newcommand{\Gabcode}[2]{\ensuremath{\mathcal{G}(#1,#2)}}
\newcommand{\IntGabcode}[1]{\ensuremath{\mathcal{IG}(#1)}}

\newcommand{\g}{\vec{g}}
\newcommand{\m}{\vec{m}}
\renewcommand{\c}{\vec{c}}

\newcommand{\M}{\vec{M}}

\IEEEoverridecommandlockouts

\begin{document}

\title{Repairing the Faure--Loidreau Public-Key Cryptosystem}
\author{\IEEEauthorblockN{Antonia Wachter-Zeh, Sven Puchinger, Julian Renner}
\IEEEauthorblockA{
Institute for Communications Engineering, Technical University of Munich (TUM), Germany\\
Email: \{antonia.wachter-zeh, sven.puchinger, julian.renner\}@tum.de
\thanks{
  J.~Renner's and A.~Wachter-Zeh's work was supported by the TUM---Institute for Advanced Study, funded by the German Excellence Initiative and European Union Seventh Framework Programme under Grant Agreement No.~291763. This work was done while S.~Puchinger was with Ulm University.
}
}}

\maketitle

\begin{abstract}
A repair of the Faure--Loidreau (FL) public-key code-based cryptosystem is proposed.
The FL cryptosystem is based on the hardness of list decoding Gabidulin codes which are special rank-metric codes.
We prove that the recent structural attack on the system by Gaborit \emph{et al.}\ is equivalent to decoding an interleaved Gabidulin code.
Since all known polynomial-time decoders for these codes fail for a large constructive class of error patterns, we are able to construct public keys that resist the attack.
It is also shown that all other known attacks fail for our repair and parameter choices.
Compared to other code-based cryptosystems, we obtain significantly smaller key sizes for the same security level.
\end{abstract}


\section{Introduction}

Public-key cryptography is the foundation for establishing secure communication between multiple parties.
Traditional public-key algorithms such as RSA are based on the hardness of factoring large numbers or the discrete logarithm problem, but can be attacked in polynomial time once a capable quantum computer exists. Code-based public-key cryptosystems are considered to be post-quantum secure, but compared to RSA their main drawback are significantly larger key sizes.

The Faure--Loidreau (FL) code-based cryptosystem~\cite{faure2006new,LoidreauHabitilation-RankMetric_2007} is based on the problem of reconstructing linearized polynomials and can be seen as linearized equivalent of the (broken) Augot--Finiasz cryptosystem \cite{AugotFiniasz-PKC-PolyReconstruction_2003}.
While the Augot--Finiasz cryptosystem is closely connected to (list) decoding Reed--Solomon codes, the 
FL cryptosystem is connected to (list) decoding Gabidulin codes, a special class of rank-metric codes~\cite{Gabidulin_TheoryOfCodes_1985}.

In contrast to McEliece or Niederreiter-type cryptosystems, where the public key is a \emph{matrix}, in the FL system, the key is only a \emph{vector}, resulting in a much smaller key.
At the time when the FL cryptosystem was designed, it was only \emph{conjectured} that Gabidulin codes cannot be list decoded efficiently. As this was \emph{proven} recently for many families of Gabidulin codes \cite{Wachterzeh_BoundsListDecodingRankMetric_IEEE-IT_2013,RavivWachterzeh_GabidulinBounds_journal}, the FL system is a very promising post-quantum secure public-key cryptosystem.

However, there are attacks on the FL cryptosystem: syndrome decoding \cite{Gaborit_DecodingAttack_2016}, an Overbeck-like attack~\cite{Overbeck-StructuralAttackGPT} which can be avoided by choosing the parameters in a certain way (cf.~\cite{LoidreauHabitilation-RankMetric_2007}) and, more severe, the recent attack by Gaborit \emph{et al.} \cite{Gaborit-KeyRecoveryFaureLoidreau} which leaves no secure set of parameters of the system.

In this paper, we propose a repair of the FL cryptosystem and prove that the attack from \cite{Gaborit-KeyRecoveryFaureLoidreau} cannot succeed anymore.
Our fundamental observation is that the attack by Gaborit \emph{et al.}~\cite{Gaborit-KeyRecoveryFaureLoidreau} on the public key is equivalent to decoding the public key as an \emph{interleaved Gabidulin code} to obtain the private key.
By choosing the public key in a way that the corresponding interleaved decoder is guaranteed to fail, the system is secured against attacks as \cite{Gaborit-KeyRecoveryFaureLoidreau}.

\section{Preliminaries}
\subsection{Notations}
Let $q$ be a power of a prime and let
$\Fq$ denote the finite field of order $q$ and $\Fqm$ its extension field of order $q^m$. 
We use $\Fq^{m \times n}$ to denote the set of all $m\times n$ matrices over $\Fq$ and $\Fqm^n =\Fqm^{1 \times n}$ for the set of all row vectors of length $n$ over $\Fqm$. 
Further, we use another field extension $\mathbb{F}_{q^{mu}}$ with $u>1$.
Thus, $\Fq \subset \Fqm \subset \Fqmu$.

Rows and columns of $m\times n$-matrices are indexed by $1,\dots, m$ and $1,\dots, n$. Denote the set of integers $[a,b] = \{i: a \leq i \leq b\}$. By $\rank_q(\vec{A})$ and $\rank_{q^m}(\vec{A})$, we denote the rank of a matrix $\vec{A}$ over $\Fq$, respectively $\Fqm$. By $ \Tr_{q^{mu}/q^m}(a)$ denote the Trace operator from an element $a \in \Fqmu$ to its projection in~$\Fqm$.

Let $(\gamma_1,\gamma_2,\dots,\gamma_{u})$ be an ordered basis of $\Fqmu$ over $\Fqm$ and $(\gamma_1^*,\gamma_2^*,\dots,\gamma_{u}^*)$ its dual basis.
By utilizing the vector space isomorphism $\Fqmu \cong \Fqm^u$, we can relate vectors $\mathbf a \in \Fqmu^n$ to matrices $\mathbf A \in \Fqm^{u \times n}$. 
Similarly, when we consider a basis of $\Fqm$ over $\Fq$, we can represent each vector of length~$n$ in $\Fqm$ as an $m \times n$ matrix in \Fq.

Denote by $\Mooremat{s}{q}{\vec{a}} \in \Fqm^{s \times n}$ the $s \times n$ Moore matrix for a vector $\vec{a} = (a_1,a_2,\dots,a_n) \in \Fqm^n$ of length $n$, i.e.:
\begin{equation*}
\Mooremat{s}{q}{\vec{a}} = \MoormatExplicit{a}{s}{n}.
\end{equation*}
If $a_1, a_2,\dots$, $a_{n}\in \Fqm$ are linearly independent over $\Fq$, then $\rank_{q^m}(\Mooremat{s}{q}{\vec{a}})=\min\{s,n\}$, cf.~\cite[Lemma 3.15]{Lidl-Niederreiter:FF1996}.

\subsection{Rank-Metric Codes and Gabidulin Codes}

The rank distance between \vec{a} and \vec{b} is the rank of the difference of the two matrix representations:
\begin{equation*}
d_{\textup{R}}(\vec{a},\vec{b})\triangleq \rank_q(\vec{a}-\vec{b}) = \rank_q(\vec{A}-\vec{B}).
\end{equation*}
An $\codelinearRank{n,k,d}$ code \mycode{C} over $\Fqm$ is a linear rank-metric code, i.e., it is a linear subspace of $\Fqm^n$ of dimension $k$ and minimum rank distance $d$.
For linear codes with $n \leq m$, the Singleton-like upper bound \cite{Delsarte_1978,Gabidulin_TheoryOfCodes_1985} implies that $d \leq n-k+1$.
If $d=n-k+1$, the code is called a \emph{maximum rank distance} (MRD) code.

Gabidulin codes \cite{Gabidulin_TheoryOfCodes_1985} are a special class of rank-metric codes and can be defined by its generator matrix as follows: 
	A linear $\Gabcode{n}{k}$ code over $\Fqm$ of length $n \leq m$ 
	and dimension $k$ is defined by its $k \times n$ generator matrix $\mathbf G_{\mycode{G}}$:
	\begin{equation*}
	\mathbf G_{\mycode{G}} = \Mooremat{k}{q}{g_1,g_2,\dots,g_n}
	\end{equation*}
	where $\vec{g}=(g_1,g_2, \dots, g_{n}) \in \Fqm^n$ has $\rank_q(\vec{g}) = n$. 
Gabidulin codes are MRD codes, i.e., $d=n-k+1$, cf.~\cite{Gabidulin_TheoryOfCodes_1985}.

\subsection{Interleaved Gabidulin Codes}
A linear (vertically, homogeneous) interleaved Gabidulin code $\IntGabcode{u;n,k}$ over $\Fqm$ of length $n \leq m$, dimension $k \leq n$, and interleaving order $u$ is defined by
\begin{equation*}
\IntGabcode{u;n,k} \triangleq
\left\lbrace
\begin{pmatrix}
\vec{c}_1\\
\vec{c}_2\\
\vdots\\
\vec{c}_u
\end{pmatrix}
: \vec{c}_i \in \Gabcode{n}{k} , \forall i \in [1,u]
\right\rbrace.
\end{equation*}
When considering random errors of rank weight $t$, the code $\IntGabcode{u;n,k}$ can be decoded uniquely with high probability up to 
$t \leq \lfloor \frac{u}{u+1}(n-k)\rfloor$ errors, cf.~\cite{Loidreau_Overbeck_Interleaved_2006,Sidorenko2011SkewFeedback,WachterzehZeh-ListUniqueErrorErasureInterpolationInterleavedGabidulin_DCC2014}.
However, it is well-known that there are many error patterns for which the known polynomial-time decoders fail.
In fact, we can explicitly construct a large class of such errors, see Lemma~\ref{lem:interleaved-fail} in Section~\ref{subsec:int-attack}.

\section{The Original Faure--Loidreau System}
\subsection{Parameters}
Let $k,n,w,u$ be positive integers such that $u<k<n$ and
\begin{equation*}
n-k>w>\nkhalffrac.
\end{equation*}
\subsection{Key Generation}
\begin{enumerate}
	\item Choose $\vec{g} \in \Fqm^n$ at random with $\rank_q(\vec{g}) = n$ and let $\mathbf G_{\mycode{G}} = \Mooremat{k}{q}{\vec{g}}$. Note that $\mathbf G_{\mycode{G}}$ is a generator matrix of a $\Gabcode{n}{k}$ code.
	\item Choose $\vec{x} \in \Fqmu^k$ at random such that $\{x_{k-u+1},\dots,x_k\}$ forms a basis of $\Fqmu$ over \Fqm.
	\item Choose $\vec{s} \in \Fqmu^w$ with $\rank_q(\vec{s}) = w$ and an invertible $\vec{P} \in \Fq^{n\times n}$ at random. Compute:
	\begin{equation*}
	\vec{z} = (\vec{s} \ | \ \0) \cdot \vec{P}^{-1}.
	\end{equation*}
\end{enumerate}

The \emph{private key} is $(\vec{x}, \vec{z})$ (or basically $\vec{z}$ since~$\vec{x}$ can be deduced from it) and the \emph{public key} is $(\vec{g}, k, \kpub, \tpub)$ where
\begin{equation*}
\kpub = \vec{x} \cdot \mathbf G_{\mycode{G}} + \vec{z},
\end{equation*}
and
\begin{equation*}
\tpub = \left\lfloor \frac{n-w-k}{2} \right \rfloor.
\end{equation*}

\subsection{Encryption}
Let $\vec{m} = (m_1,\dots,m_{k-u},0,\dots,0) \in \Fqm^k$ be the plaintext. Note that the upper $u$ coefficients are chosen to be zero in order to decrypt $\alpha$. 
The encryption of $\vec{m}$ works as follows:
\begin{enumerate}
	\item Choose $\alpha \in \Fqmu$ at random. 
	\item Choose $\vec{e} \in \Fqm^n$ such that $\rank_q(\vec{e}) \leq \tpub$ at random.
	\item Calculate the ciphertext $\vec{c} \in \Fqm^n$:
	\begin{equation*}
		\vec{c} = \vec{m}\cdot\mathbf G_{\mycode{G}} + \Tr_{q^{mu}/q^m}(\alpha \kpub) +\vec{e}.
	\end{equation*}
\end{enumerate}

\subsection{Decryption}\label{subsec:decryption}
\begin{enumerate}
	\item Compute $\vec{c}\vec{P}$ which results in
	\begin{equation*}
	\hspace{-4ex}\vec{c}\vec{P} = (\vec{m} + \Tr_{q^{mu}/q^m}(\alpha\vec{x}))\vec{G}_{\mycode{G}}\vec{P} + (\Tr_{q^{mu}/q^m}(\alpha \vec{s}) | \vec{0}) + \vec{e}\vec{P}.
	\end{equation*}
	\item Let $\vec{G}^{\prime}$ denote the $k \times (n-w)$ matrix obtained by removing the first $w$ columns of $\vec{G}_{\mycode{G}}\vec{P}$ and let $\vec{e}^{\prime}$ and $\vec{c}^{\prime}$ denote the last $n-w$ positions of $\vec{e}\vec{P}$ and $\vec{c}\vec{P}$, respectively. Then,
	\begin{equation*}
	\vec{c}^{\prime} = (\vec{m} + \Tr_{q^{mu}/q^m}(\alpha\vec{x}))\vec{G}^{\prime} + \vec{e}^{\prime}.
	\end{equation*}
	Apply a decoding algorithm on $\vec{c}^{\prime}$. This gives $\vec{m}^{\prime} \triangleq\vec{m} + \Tr_{q^{mu}/q^m}(\alpha\vec{x})$ and is possible since $\vec{G}^{\prime}$ is a generator matrix of a $\Gabcode{n-w}{k<n-w}$ code.
	\item Since $m_i=0$ for $i=k-u+1,\dots,k$, compute $\alpha$ by:
	$\alpha = \sum_{i=k-u+1}^{k}m_i^{\prime}x_i^*$,
	where $\{x_{k-u+1}^*,\dots,x_k^*\}$ is a dual basis to $\{x_{k-u+1},\dots,x_k\}$.
	\item Calculate $\vec{m} = \vec{m}^{\prime}-\Tr_{q^{mu}/q^m}(\alpha\vec{x})$.
\end{enumerate}

\section{Public~Key~as~Interleaved Gabidulin~Code}\label{subsec:int-intgab}
Recall the definition of the public key:
\begin{equation*}
\kpub = \vec{x} \cdot \mathbf G_{\mycode{G}} + \vec{z},
\end{equation*}
where $\vec{x} \in \Fqmu^k$, $\vec{G}_{\mycode{G}} \in \Fqm^{k\times n}$ is the generator matrix of a $\Gabcode{n}{k}$ code, and $\vec{z} \in \Fqmu^n$ with $\rank_q(\vec{z})=w$.

Let $\kpub = \sum_{i=0}^{u-1} \kpub^{(i)}\gamma_i^*$, $\vec{z} = \sum_{i=0}^{u-1} \vec{z}_i\gamma_i^*$ and $\vec{x} = \sum_{i=0}^{u-1} \vec{x}_i\gamma_i^*$, where $\kpub^{(i)}, \vec{z}_i, \vec{x}_i$ have coefficients in $\Fqm$.

Then, we obtain the following representation of the public key $\kpub$ as a $u \times n$ matrix in $\Fqm$:
\begin{equation*}
\begin{pmatrix}
\kpub^{(1)}\\
\kpub^{(2)}\\
\vdots\\
\kpub^{(u)}
\end{pmatrix}
=
\begin{pmatrix}
\vec{x}_1\\
\vec{x}_2\\
\vdots\\
\vec{x}_u
\end{pmatrix}\!\cdot \vec{G}_{\mycode{G}}
+\!
\begin{pmatrix}
\vec{z}_1\\
\vec{z}_2\\
\vdots\\
\vec{z}_u
\end{pmatrix}
=
\begin{pmatrix}
\vec{x}_1\cdot \vec{G}_{\mycode{G}}\\
\vec{x}_2\cdot \vec{G}_{\mycode{G}}\\
\vdots\\
\vec{x}_u\cdot \vec{G}_{\mycode{G}}
\end{pmatrix}
+\!
\begin{pmatrix}
\vec{z}_1\\
\vec{z}_2\\
\vdots\\
\vec{z}_u
\end{pmatrix}.
\end{equation*}

Since $\vec{x}_i \cdot \vec{G}_{\mycode{G}}$ is a codeword of a $\Gabcode{n}{k}$ code, $\forall i \in [1,u]$, this matrix can be seen as a codeword from an $\IntGabcode{u;n,k}$ code, corrupted by an error of rank weight $w$.

\section{Possible Attacks on the FL System}
\subsection{(List) Decoding}
An attacker can try to (list) decode either the public key $\kpub$ or the ciphertext $\vec{c}$. However, in both cases, the number of errors that affect the Gabidulin codeword is larger than the unique decoding radius $\nkhalffrac$. Therefore, a \emph{list} decoding algorithm with larger decoding radius has to be applied. However, for many families of Gabidulin codes such an efficient list decoding algorithm cannot exist \cite{Wachterzeh_BoundsListDecodingRankMetric_IEEE-IT_2013,RavivWachterzeh_GabidulinBounds_journal} and therefore an efficient list decoding attack (on the public key or the ciphertext) does not exist.

\subsection{Syndrome Decoding Attack}
The ciphertext can be interpreted as a codeword from a code of dimension $k$ (see~\cite{faure2006new}), generated by the generator matrix
\begin{equation*}
\begin{pmatrix}
\Mooremat{k-u}{q}{\vec{g}}\\
\Tr(\gamma_1 \kpub)\\
\vdots\\
\Tr(\gamma_u \kpub)\\
\end{pmatrix}.
\end{equation*}
Since the structure of this code only permits decoding like a random rank-metric code, it can be decoded with the syndrome decoding attack from~\cite{aragon:hal-01618464} whose complexity is in the order of $(n-k)^3m^3q^{\tpub \lceil \frac{(k+1)m}{n} \rceil -m}$.

\subsection{Gaborit--Otmani--Kalachi Attack}

\begin{theorem}[GOK Attack {\cite[Thm.~1]{Gaborit-KeyRecoveryFaureLoidreau}}]\label{thm:GOK-attack-success}
	Let $\gamma_1,\dots,\gamma_u \in \Fqmu$ be a basis of $\Fqmu$ over $\Fqm$ and let $\vec{z}_i = \Tr_{q^{mu}/q^m}(\gamma_i \vec{z})$, for $i=1,\dots u$.
	
	If the matrix $\vec{Z} \in \Fqm^{u \times n}$ with $\vec{z}_1,\dots,\vec{z}_u$ as rows, satisfies
	\begin{equation*}
	\rank_{q^m} (\Mooremat{n-k-w}{q}{\vec{Z}}) = w,
	\end{equation*}
	then the private key $(\vec{x}, \vec{z})$ can be recovered from $\vec{G}_{\mycode{G}},\kpub$ with $\mathcal{O}(n^3)$ operations in \Fqmu.
\end{theorem}

\subsection{Interleaved Decoding}\label{subsec:int-attack}
Based on the interpretation of Section~\ref{subsec:int-intgab}, another possible attack is to apply a decoder for an interleaved Gabidulin code to the public key $\kpub$. 
If $w \leq  \frac{u}{u+1}(n-k)$, such a decoder will return $\vec{x}$ with high probability, but fail in certain cases.
\begin{lemma}[Interleaved Decoding~{\cite{Loidreau_Overbeck_Interleaved_2006}, \cite{Sidorenko2011SkewFeedback}, \cite[p.~64]{Wachterzeh_DecodingBlockConvolutionalRankMetric_2013}}]\label{lem:interleaved-fail}
	Let~$w$ be an integer with $w \leq \frac{u}{u+1}(n-k)$ and $\vec{c}_i=\vec{x}_i \cdot \vec{G}_{\mathcal{G}}$.
	All efficient decoders for $\IntGabcode{u;n,k}$ codes fail to correct an error $\vec{z}\in \Fqmu^n$ with $\vec{z} = \sum_{i=0}^{u-1} \vec{z}_i\gamma_i^*$ and $\rank_q(\vec{z})=w$ if 
\begin{equation}
\rank_{q^m}
\begin{pmatrix}
\Mooremat{n-w-1}{q}{\vec{g}}\\
\Mooremat{n-k-w}{q}{\vec{c}_1+\vec{z}_1}\\
\Mooremat{n-k-w}{q}{\vec{c}_2+\vec{z}_2}\\
\vdots\\
\Mooremat{n-k-w}{q}{\vec{c}_u+\vec{z}_u}\\
\end{pmatrix}
< n-1
\end{equation}
\end{lemma}
Since $\rank_{q^m}(\Mooremat{n-w-1}{q}{\vec{g}})=n-w-1$, the interleaved decoder fails if
\begin{equation}\label{eq:fail-interleaved}
\rank _{q^m} 
\begin{pmatrix}
\Mooremat{n-k-w}{q}{\vec{z}_1}\\
\Mooremat{n-k-w}{q}{\vec{z}_2}\\
\vdots\\
\Mooremat{n-k-w}{q}{\vec{z}_u}\\
\end{pmatrix}
< w.
\end{equation}

Thus, the basic idea of our repair (see~Section~\ref{sec:repair}) is to choose~$\vec{z} \in \Fqmu^n$ such that~\eqref{eq:fail-interleaved} holds.
This makes it impossible to attack the system by decoding the public key $\kpub$ with an interleaved Gabidulin decoder. To our knowledge, this connection has not been known before.

\subsection{Equivalence of Attack from \cite{Gaborit-KeyRecoveryFaureLoidreau} and Interleaved Decoding}

\begin{theorem}\label{thm:equiv-int-gok}
	The attack from \cite{Gaborit-KeyRecoveryFaureLoidreau} fails if and only if the attack based on interleaved decoding in Section~\ref{subsec:int-attack} fails. In particular, it fails if \eqref{eq:fail-interleaved} holds.
\end{theorem}
\begin{IEEEproof}
	By definition $\vec{z}_i=\Tr_{q^{mu}/q^m}(\gamma_i \vec{z}) \in \Fqm^n$, i.e., $\vec{z} = \sum_{i=0}^{u-1} \vec{v}_i\gamma_i^*$, where $(\gamma_1^*,\gamma_2^*,\dots,\gamma_{u}^*)$ denotes a dual basis to $(\gamma_1,\gamma_2,\dots,\gamma_{u})$.
The matrix $\Mooremat{n-w-k}{q}{\vec{Z}}$ from Theorem~\ref{thm:GOK-attack-success} is row-space equivalent to:
\begin{align*}
\begin{pmatrix}
\Mooremat{n-k-w}{q}{\vec{z}_1}\\
\Mooremat{n-k-w}{q}{\vec{z}_2}\\
\vdots\\
\Mooremat{n-k-w}{q}{\vec{z}_u}\\
\end{pmatrix}.
\end{align*}	
The rank of this matrix cannot be come larger than~$w$ (since any vector in the right kernel of this matrix has rank weight at least $n-w$ \cite[Algorithm 3.2.1]{Overbeck_Diss_InterleveadGab}). 
Thus, the failures of Theorem~\ref{thm:GOK-attack-success} and Lemma~\ref{lem:interleaved-fail},~\eqref{eq:fail-interleaved} are equivalent.
\end{IEEEproof}

\subsection{Linearization Attack}

\begin{theorem}[Linearization Attack {\cite{faure2006new}}]
Let $\kpub^{(i)} = \Tr_{q^{mu}/q^m}(\gamma_i \kpub)$ for $i=1,\dots,u$ and
\begin{align}
\M =
\begin{pmatrix}
\Mooremat{\tpub+1}{q}{\c} \\
-\Mooremat{\tpub+1}{q}{\kpub^{(1)}} \\
\vdots \\
-\Mooremat{\tpub+1}{q}{\kpub^{(u)}} \\
-\Mooremat{k+\tpub-u}{q}{\g}
\end{pmatrix}. \label{eq:M_linearization_attack}
\end{align}
Then, the encrypted message $\m$ can be efficiently recovered if the left kernel of $\M$ has dimension $\dim(\ker(\M)) = 1$.
\end{theorem}

If $(u+2)\tpub + k > n$, then $\M$ has at least two more rows than columns and we have $\dim(\ker(\M))>1$.
If $\kpub$ is random and $(u+2)\tpub + k \leq n$, the attack is efficient with high probability~\cite{faure2006new}.
This implies the following.
\begin{corollary}
The linearization attack in \cite{faure2006new} is inefficient if
$\tpub > \tfrac{n-k}{u+2}$
{and its work factor is $q^{m(\tpub(u+2)-n+k+1)}$}.
In this case, we must choose
\begin{equation*}
w < n-k-2\tpub < \tfrac{u}{u+2} (n-k).
\end{equation*}
\end{corollary}

\subsection{Algebraic Attacks}

Faure and Loidreau \cite{faure2006new} also described two message attacks of exponential worst-case complexity. The first one is based on computing gcds of polynomials of degrees
\begin{equation}
q^{m(u-1)}\frac{q^{m (\tpub+1)}-1}{q^m-1}. \label{eq:algebraic_attack_work_factor}
\end{equation}
Since computing the gcd of two polynomials can be implemented in quasi-linear time in the polynomials' degree,~\eqref{eq:algebraic_attack_work_factor} gives an estimate on the work factor of this attack.
The second algebraic attack is based on finding Gr\"obner bases of a system of
$n_\mathrm{p} = \tbinom{n}{k+2 \tpub-u+1}$ many polynomials of degree approximately $d_\mathrm{p} = \tfrac{q^{\tpub+1}-1}{q-1}$.
The attack is only efficient for small code parameters, cf.~\cite[Sec.~5.3]{faure2006new}. Since the average-case complexity of Gr\"obner bases algorithms is hard to estimate, we cannot directly relate $n_\mathrm{p}$ and $d_\mathrm{p}$ to the attack's work factor. Faure and Loidreau choose the code parameters such that $n_\mathrm{p} \approx 2^{32}$ and $d_\mathrm{p} = 127$ and claim that the attack is inefficient for these values. Our example parameters in Section~\ref{sec:parameters} result in at least these values.

\subsection{Overbeck-like Attack}

The key attack described in \cite[Ch.~7, Sec.~2.1]{LoidreauHabitilation-RankMetric_2007} is based on a similar principle as Overbeck uses to attack the McEliece cryptosystem based on rank-metric codes~\cite{Overbeck-StructuralAttackGPT}.
The attack from \cite[Ch.~7, Sec.~2.1]{LoidreauHabitilation-RankMetric_2007} cannot be applied if
$w \geq n-k -\frac{k-u}{u-1}$.

\section{Our Repair}\label{sec:repair}

We choose $n-k -\frac{k-u}{u-1} \leq w<n-k$ (i.e., the Overbeck-like attack is inefficient). Notice that this imposes a rate restriction, e.g., for $u=3$, we get that $ R \gtrsim
\frac{7}{11}$ which does not cause any problem in a code-based cryptosystem. 

Our repair is based on choosing $\vec{z} = \sum_{i=0}^{u-1} \vec{z}_i\gamma^*_i$ in a way that
\begin{equation}\label{eq:matrix-repair}
\rank _{q^m} 
\begin{pmatrix}
\Mooremat{n-k-w}{q}{\vec{z}_1}\\
\Mooremat{n-k-w}{q}{\vec{z}_2}\\
\vdots\\
\Mooremat{n-k-w}{q}{\vec{z}_u}\\
\end{pmatrix}
\triangleq \RkError
< w.
\end{equation}
In this case, an interleaved decoder, see~\eqref{eq:fail-interleaved}, and therefore also the attack by Gaborit \emph{et al.} \cite{Gaborit-KeyRecoveryFaureLoidreau} fails, see Theorem~\ref{thm:equiv-int-gok}.

One possible choice is to set $\vec{z}_1 = \vec{z}_2 = \dots = \vec{z}_u$ with $\rank_q(\vec{z}_1)= \rank_q(\vec{z}) = w$. In this case $\RkError = \rank _{q^m} (\Mooremat{n-k-w}{q}{\vec{z}_1}) = n-k-w < w$. This is true since $w>\nkhalffrac$, i.e., $\Mooremat{n-k-w}{q}{\vec{z}_1}$ has more linearly independent elements in the first row than rows. Since  $w>\nkhalffrac$, the rank is less than $w$ and an interleaved decoder fails. 

In the next section, we will show that the restriction on $\vec{z}$ from~\eqref{eq:matrix-repair} does not pose a problem in terms of the security level. In particular, choosing $\vec{z}_1 = \vec{z}_2 = \dots = \vec{z}_u$ with $\rank_q(\vec{z}_1) = w$ provides the largest security level among all choices for~\eqref{eq:matrix-repair} and provides an explicit construction of $\vec{z}$. 

We therefore propose the following modification of the key generation algorithm which results in $\vec{z}_1 = \vec{z}_2 = \dots = \vec{z}_u$ with $\rank_q(\vec{z}_1)= \rank_q(\vec{z}) = w$.\\

\textbf{Repair:} 
\textit{
Replace Step~3 in \textbf{Key Generation} by:\\
3.) Choose $\vec{s}_1 \in \Fqm^w$ at random with 
$\rank_q(\vec{s}_1) = w$, let $\vec{s} = \vec{s}_1 \cdot\sum_{i=0}^{u-1}\gamma_i^* \in \Fqmu^w$ and let $\vec{P} \in \Fq^{n\times n}$ be invertible. Compute:
\begin{equation}\label{eq:repair-z}
\vec{z} = (\vec{s} \ | \ \0) \cdot \vec{P}^{-1}.
\end{equation}
All other steps of the algorithms remain the same.}\\

Clearly, we restrict the choice of $\vec{z}$ in Step~3 of the Key Generation algorithm.
We will see that there are still enough possibilities for $\vec{z}$ to preserve a high security level.

\section{Security Analysis}

Our repair is designed such that an interleaved decoding attack and the attack from \cite{Gaborit-KeyRecoveryFaureLoidreau} do not work. In this section, we analyze the security level resulting from (A.) brute-forcing~$\vec{z}$, (B.) searching the kernel  of~\eqref{eq:matrix-repair}, and (C.) the linearization attack.

Note that guessing $\alpha \in \Fqmu$ correctly makes it possible to reconstruct the secret message.
Thus, the security of the system is limited by the number of possible values of~$\alpha$ which is $q^{mu}$.
\subsection{Number of Possible Vectors $\vec{z}$}
	For the explicit repair of~\eqref{eq:repair-z} (where $\RkError=n-k-w$), the number of matrices $\vec{s}_1 \in \Fqm^w$ with 
	$\rank_q(\vec{s}_1) = w$ is larger than $0.288 \cdot q^{mw}$, see \cite[Lemma~3.13]{Overbeck_Diss_InterleveadGab} and the number of full-rank matrices $\vec{P} \in \Fq^{n\times n}$ is larger than $0.288 \cdot q^{n^2}$. Thus, the number of possible vectors~$\vec{z}$ in~\eqref{eq:repair-z} is larger than:
	\begin{equation*}
	0.288^2 \cdot q^{mw + n^2}.
	\end{equation*}
	Since $u < n \leq m$, this is usually larger than $q^{mu}$ and trying all possible~$\vec{z}$ does not reduce the security of the system.

	In the general case of~\eqref{eq:matrix-repair} (i.e., any $n-k-w \leq \RkError<w$), the number of Moore matrices as in~\eqref{eq:matrix-repair} and such that $\rank_q(\vec{z})=w$ is given from the \emph{failure probability} of an interleaved Gabidulin decoder (cf.~\cite{Sidorenko2011SkewFeedback}). The number of such matrices therefore equals
	\begin{equation*}
	|\vec{z} \in \Fqmu^n \, : \ \rank_q(\vec{z})=w|\cdot P_f > 0.288^2\cdot q^{muw+n^2}\cdot P_f.
	\end{equation*}
	However, a \emph{lower} bound on the failure probability is not known.
	As an approximation, we can use $P_f \approx 1/(q^m)$ and therefore the number of such matrices is larger than $0.288^2\cdot q^{uw+n/m}$. This, and the next subsection, make clear that~\eqref{eq:repair-z} and $\RkError=n-k-w$ is the most secure way to realize~\eqref{eq:matrix-repair}.

\subsection{Size of the Solution Space of~\eqref{eq:matrix-repair}}
Finding the kernel of the matrix in~\eqref{eq:matrix-repair} with an interleaved decoder is equivalent solving a linear system of equations based on the syndromes with $w$ unknowns and $\RkError$ linearly independent equations, cf.~\cite[Section~4.1]{Wachterzeh_DecodingBlockConvolutionalRankMetric_2013}.
The size of the solution space of this decoder is
$(q^m)^{w-\RkError}$,
which is maximized for the smallest-possible value of $\RkError$, i.e., $\RkError = n-k-w$. In this case the size of the solution space is
$(q^m)^{2w-(n-k)}$.
For $w = \lfloor\frac{u}{u+1}(n-k)\rfloor$, we get:
\begin{equation*}
q^{m \left(2\lfloor\frac{u}{u+1}(n-k)\rfloor-(n-k)\right)} \approx q^{m(n-k)}.
\end{equation*}
Since the size of the solution space is maximal for $\RkError = n-k-w$, the explicit repair from~\eqref{eq:repair-z} is the most secure choice {in this sense}.

\begin{table*}[!t]
	\caption{Comparison of the McEliece (based on Goppa codes), the Loidreau, our repaired FL cryptosystem, and the QC-MDPC scheme}
	\begin{center}
		\begin{tabular}{l|c|c|c|c|c|c|c|c|c||c|c|r }
			Method &$q$ & $u$ & $k$ & $n$ & $m$ & $w$  & $\tau$ & $t_{\text{Loi}}$ & $\lambda$ & Security level & Rate & Key size  \\
			\hline  \hline
			McEliece &2 & & 1436 & 1876 & 11 & &  41 & & & 80.04 & 0.77 & 78.98 KB \\
			\hline
			Loidreau &2 & & 32 & 50 & 50 &  & & 3 &3  & 80.93 & 0.64 & 3.60 KB \\
			\hline
                  Repaired FL &2 & 3 & 31 & 61 & 61 & 16  & & & & 90.00 & 0.46 & 1.86 KB \\
                  \hline
                                          QC-MDPC & 2  &  &  4801 & 9602 & & & & &  & 80.00 & 0.50 & 0.60 KB \\
                  \hline \hline
			McEliece &2 & & 2482 & 3262 & 12 &  & 66 & & & 128.02 & 0.76  & 242.00 KB  \\
			\hline
			Loidreau &2 & & 40 & 64 & 96 &  & & 4 &3 & 139.75 & 0.63 & 11.52 KB \\
			\hline
                  Repaired FL&2 & 3 & 31 & 63 & 63 & 18  & & & & 141.56 & 0.44 & 1.98 KB \\
                  \hline
                                    QC-MDPC & 2  &  &  9857 & 19714 & & & & &  & 128.00 & 0.50 & 1.23 KB \\
			\hline \hline
			McEliece&2 & & 5318 & 7008 & 13 &  & 133 & && 257.47 & 0.76  & 1123.43 KB  \\
			\hline
			Loidreau&2 & &80 &120  &128  &  &  & 4  &5 &261.00  & 0.67  & 51.20 KB  \\
			\hline
                  Repaired FL&2 & 4  & 48 & 82 & 82 & 20  & & & & 262.35 & 0.54 & 4.20 KB \\
                  \hline
                                    QC-MDPC & 2  & & 32771  &  65542  & & & & &  & 256.00 & 0.50 & 4.10 KB \\
		\end{tabular}
	\end{center}
	\label{tab:security_para}
	\hrulefill
\end{table*}

\subsection{Linearization Attack}

\begin{theorem}
Let $\M$ be as in \eqref{eq:M_linearization_attack}. Then,
$\rank_{q^m}(\M) \leq \varphi + k + 2\tpub -u$.
\end{theorem}

\begin{IEEEproof}
We can write
\begin{align*}
\kpub^{(i)} &= \Tr_{q^{mu}/q^m}(\gamma_i \kpub) \\
&= \Tr_{q^{mu}/q^m}(\gamma_i \vec{x}) \cdot \Mooremat{k}{q}{\g} + \vec{z}_i,
\end{align*}
so by elementary row operations, we can transform $\M$ into
\begin{align*}
\M' = \begin{pmatrix}
\Mooremat{\tpub+1}{q}{\c} \\
-\Mooremat{\tpub+1}{q}{\vec{z}_1} \\
\vdots \\
-\Mooremat{\tpub+1}{q}{\vec{z}_u} \\
-\Mooremat{k+\tpub-u}{q}{\g}
\end{pmatrix}.
\end{align*}
Due to $w+2\tpub<n-k$, the matrix $\Mooremat{\tpub+1}{q}{\vec{z}_i}$ is a sub-matrix of $\Mooremat{n-k-w}{q}{\vec{z}_i}$, so
\begin{align*}
&\rank_{q^m} (\M) = \rank_{q^m} (\M') \\
&\leq \varphi + \rank_{q^m} (\Mooremat{\tpub+1}{q}{\c}) + \rank_{q^m} (\Mooremat{k+\tpub-u}{q}{\g}) \\
&= \varphi + k + 2\tpub -u.
\end{align*}

\vspace{-4ex}
\end{IEEEproof}

The linearization attack is inefficient if the rank of $\M$ is smaller than its number of rows, which implies the following.

\begin{corollary}
If $\varphi < u (\tpub+1)$ the linearization attack in \cite{faure2006new} is inefficient.
This is fulfilled for 
$\tpub > \tfrac{n-k-w}{u}+1 \geq \tfrac{\varphi}{u}+1$,
and there are such values of $\tpub<n-k-2w$ for any $u>1$.
\end{corollary}

\subsection{Attack By Moving to Another Close Error}
The following attack \cite{Johan_attack} tries to move the vector $\vec{z}$ (which we have chosen such that the interleaved decoder fails) on a close vector for which the interleaved decoder for $\kpub$ does not fail. 
Therefore, a vector $\vec{y} \in \Fqm^{u \times n}$ is needed such that for $\vec{z}^\prime\triangleq\vec{z} + \vec{y}$ it holds that $\rank_q(\vec{z}^\prime) \leq w$ and that the rank of the matrix from~\eqref{eq:matrix-repair} over $\Fqm$ is at least $w$.

Nielsen suggested to find such a vector by guessing $2w-n+k$ independent vectors from $\Fq^n$ which are in the $\Fq$-row space of $\vec{z}$, put them as the first rows of a matrix in $\Fq^{um\times n}$ (the remaining rows are zeros) and use its mapping to a matrix in $\Fqm^{u \times n}$ as matrix $\vec{y}$.
That way, $\vec{z}^\prime$ is in the row space of $\vec{z}$ and $\rank_q(\vec{z}^\prime) \leq w$ is guaranteed. Further, the rank of the matrix from~\eqref{eq:matrix-repair} over $\Fqm$ is increased to $w$ with high probability.

The complexity of this attack is dominated by the complexity of finding $2w-n+k$ independent vectors from $\Fq^n$ which are in the $\Fq$-row space of $\vec{z}$, i.e.:
\begin{equation*}
\text{WF}_{\text{Err}} = q^{(2w-n+k)(n-w)}.
\end{equation*}

\section{Parameters and Key Sizes}\label{sec:parameters}

To evaluate the performance of the repaired FL cryptosystem, it is compared with McEliece's cryptosystem based on Goppa codes using list decoding \cite{Barbier_2011}, Loidreau's new rank-metric code-based encryption scheme~\cite{Loidreau-GPT-ACCT2016,Loidreau2017-NewRankMetricBased} and the QC-MDPC cryptosystem \cite{Misoczki_2013}.
The most efficient attack on McEliece has work factor (cf.~\cite{Barbier_2011})\\[-4ex]

\begin{small}
	\begin{equation*}
	\text{WF}_{\text{ME}}= \text{min} \Bigg\{ \frac{1}{2} \binom{n}{\tau} \binom{n-k}{\tau-p}^{-1}\hspace{-.8ex}
	\binom{k}{p}^{-1/2}\hspace{-2ex}:0\leq p \leq \text{min}\{\tau,k\} \Bigg\}
	\end{equation*}
\end{small}

\hspace{-2.3ex}operations, where $\tau$ is the binary Johnson bound. 

The work factor of Loidreau's system~\cite{Loidreau-GPT-ACCT2016,Loidreau2017-NewRankMetricBased} is
\begin{equation*}
 \text{WF}_{\text{Loi}}= m^3 q^{(t_{\text{Loi}}-1) \lfloor(k\cdot \text{min}(m,n))/n \rfloor },
\end{equation*}
operations, where $t_{\text{Loi}} \cdot \lambda = \nkhalffrac$.

By choosing
\begin{align*}
2 \leq u < &\, k < n \leq m,\\
 \bigg \lfloor \frac{n-k}{2} \bigg \rfloor < & \, w < \frac{u}{u+2} (n-k),\\
  n-k-\frac{k-u}{u-1} \leq & \, w
\end{align*}
the work factors of the attacks on the repaired FL can be summarized as follows. There are
$\text{WF}_{\alpha}=  q^{mu}$
choices of $\alpha$, the syndrome decoding attack requires
\begin{equation*}
\text{WF}_{\text{Dec}} = (n-k)^3m^3q^{\tpub \lceil \frac{(k+1)m}{n} \rceil -m}
  \end{equation*}
operations \cite{aragon:hal-01618464}, the linearization attack has a work factor of
\begin{equation*}
  \text{WF}_{\text{Lin}} = q^{m(ut_{\text{pub}} + u + 1 - \varphi)},
\end{equation*}
there are
$\text{WF}_{\boldsymbol{z}} = 0.288^2  q^{mw+n^2}$
  possible vectors $\boldsymbol{z}$, the interleaved decoding attack needs
$\text{WF}_{\text{ILD}} = q^{m(w-\varphi)}$
operations, the first algebraic attack has a work factor $\text{WF}_{\text{Alg}} = q^{m(u-1)}\frac{q^{m (\tpub+1)}-1}{q^m-1}$ and the Overbeck-like attack has complexity at least $\text{WF}_{\text{ILD}}$. 

Table \ref{tab:security_para} proposes parameters for expected work factors of around $2^{80}$, $2^{130}$ and $2^{260}$. The work factor of the repaired FL system stems from the number of operations required by the most efficient attack which is the attack by moving to another close error for $2^{80}$ and the algebraic attack for $2^{130}$ and $2^{260}$. We observe that in all cases McEliece has the highest rate followed by Loidreau, FL and QC-MDPC. The results show further that FL and QC-MDPC require much smaller key sizes compared to Loidreau and McEliece.
Since public-key cryptosystems are mostly used for encrypting small data packages (usually they are used to exchange the private key of a symmetric cryptosystem), small key sizes are more important than high code rates.
The QC-MDPC scheme gives no guarantee that the ciphertext can be decrypted as decoding these codes might fail
while the repaired FL system guarantees decryption.
Hence, the repaired FL cryptosystem has advantages compared to the other mentioned systems and should be considered as an alternative of small key size.

\bibliographystyle{IEEEtran}
\bibliography{main}

\end{document}